\documentclass[a4paper,
               keeplastbox,   
               ]{jacow}
\usepackage{pdfpages,multirow,ragged2e} %
\makeatletter%
	\ifboolexpr{bool{xetex}}
	 {\renewcommand{\Gin@extensions}{.pdf,%
	                    .png,.jpg,.bmp,.pict,.tif,.psd,.mac,.sga,.tga,.gif,%
	                    .eps,.ps,%
	                    }}{}
\makeatother

\ifboolexpr{bool{xetex} or bool{luatex}} 
 {}                                      
 {\usepackage[utf8]{inputenc}}           

\usepackage[USenglish]{babel}

\ifboolexpr{bool{jacowbiblatex}}%
 {%
  \addbibresource{jacow-test.bib}
  \addbibresource{biblatex-examples.bib}
 }{}
\begin{document}

\title{\textit{Ab initio} study of antisite defects in Nb$_3$Sn: phase diagram and impact on superconductivity\thanks{This work was supported by the US National Science Foundation under award PHY-1549132, the Center for Bright Beams.}}

\author{N. S. Sitaraman\thanks{nss87@cornell.edu}, T. A. Arias, R. D. Porter, \\
M. U. Liepe, Cornell University Department of Physics, 14850, Ithaca, USA \\
J. Carlson, A. R. Pack, M. K. Transtrum, \\
Brigham Young University Department of Physics, 84602, Provo, USA}

\maketitle

\begin{abstract}
Antisite defects play a critical role in Nb$_3$Sn superconducting radio frequency (SRF) cavity physics. Such defects are the primary form of disorder in Nb$_3$Sn, and are responsible for stoichiometry variations, including experimentally observed tin-depleted regions within grains and tin-rich regions around grain boundaries. But why they cluster to form regions of different stoichiometries and how they affect the SRF properties of Nb$_3$Sn cavities are not fully understood. Using \emph{ab initio} techniques, we calculate the A15 region of the Nb-Sn phase diagram, discuss a possible modification to the phase diagram near grain boundaries, and calculate $T_c$ as a function of stoichiometry, including experimentally inaccessible tin-rich stoichiometry. We find that the impact of antisite defects on the density of states near the Fermi-level of Nb$_3$Sn plays a key role in determining many of their properties. These results improve our understanding of the obstacles facing Nb$_3$Sn SRF systems, and how modified growth processes might overcome them.
\end{abstract}

\maketitle

\section{INTRODUCTION}

Over 40 years after the first Nb$_3$Sn cavities were made, Nb$_3$Sn finally appears poised to overtake niobium as material of choice for SRF applications. With the potential to operate at a higher temperature, higher quality factor, and higher accelerating gradient than niobium, Nb$_3$Sn could revolutionize the field, making new applications possible and making existing applications more economical and more compact. However, progress towards the pristine, uniform Nb$_3$Sn surfaces necessary to realize this potential has been intermittent due to a poor understanding of what defects affect SRF performance and how those defects form during Nb$_3$Sn layer growth \cite{Posen,Ryan,Liepe}.

One type of defect commonly observed in Nb$_3$Sn samples is the tin-depleted region. Stoichiometry studies of Nb$_3$Sn layers grown with the modern vapor deposition procedure show that while most of the material has stoichiometry very near the ideal 25\% Sn, with $T_c$ around 18 K, significant regions on the scale of hundreds of nanometers have stoichiometry near 18\% Sn, with $T_c$ around 6 Kelvin \cite{LINAC,DHall,Lee}. This inhomogeneity does not come as a complete surprise, as it has long been understood that Nb$_3$Sn represents one possible composition of the Nb-Sn A15 phase, which can have tin content as high as about 26\%, and as low as about 18\% \cite{Charlesworth}. But theories for why tin-depleted regions form, let alone where in the layer they form, are lacking. The position of these regions relative to the surface is a matter of particular concern, as a poorly superconducting region near the surface could cause a cavity to quench.

A type of defect observed more recently is the tin-rich grain boundary. Atomic-precision microscopy studies of Nb$_3$Sn grain boundaries reveal an excess of tin which, surprisingly, extends at least a nanometer into the crystal on either side of the grain boundary core \cite{GB}. Theories for how tin-rich grain boundaries form are a little more straightforward than theories for how tin-depleted regions form, as grain boundaries are thought to be pipelines for tin diffusion during layer growth \cite{Diffusion}. However, the details of their nature are more mysterious: the Nb-Sn phase diagram does not allow for the $\sim$30\% Sn concentration observed in these regions, so it is unclear what physical mechanism stabilizes them, and what their superconducting properties might be. Furthermore, the scale of these regions is similar to the coherence length of $\sim$3~nm in Nb$_3$Sn, making the details of their structure and superconducting properties critically important \cite{Coherence}.

This paper seeks to understand how, when, and why antisite defects segregate in Nb$_3$Sn during layer growth to form tin-depleted regions and tin-rich grain boundaries, and to calculate the effect of segregated antisite defects on the superconducting properties. To do so, we use density functional theory (DFT) implemented in our in-house JDFTx software \cite{JDFTx}. Specifically, we determine the effect of antisite defects on the electronic structure of Nb$_3$Sn, and find that their effect on the density of states at the Fermi-level is of particular importance in determining how these defects form and interact during layer growth, and in determining how they affect $T_c$.

\section{METHODOLOGY}

All calculations are performed using the Perdew-Burke-Ernzerhof (PBE) version of the generalized gradient approximation with Nb $4p^{6}5s^{2}4d^{3}$ and Sn $4d^{10}5s^{2}5p^{2}$ ultrasoft pseudopotentials \cite{GGA,Psp}. For total-energy calculations involving lattice minimization, we use a 20 Hartree planewave cutoff energy in order to minimize the effect of the changing reciprocal space wave-function basis on calculated energy differences. All other calculations use a 12 Hartree plane-wave cutoff energy. All calculations used a 4.5 milliHartree electron temperature, chosen to be close to the actual experimental growth temperature for Nb$_3$Sn of about 1420 K. For zero-temperature properties, we use the cold-smearing method developed by Marzari and use the same 4.5 milliHartree smearing width in order to ensure that zero-temperature energies are converged to the same tolerance as high-temperature energies with respect to k-point sampling \cite{Cold}. For total-energy calculations on the A15 phase, we use a 64-atom unit cell, and a 4$\times$4$\times$4 gamma-centered k-point mesh to sample the first Brillouin zone. For the Nb$_6$Sn$_5$ phase, we use a 44-atom unit cell, and a 3$\times$6$\times$9 k-point mesh. For the bcc phase, we use a 54-atom unit cell, and a 5$\times$5$\times$5 k-point mesh.

To calculate T$_c$ as a function of stoichiometry, we consider \emph{all possible} unique antisite defects in (a) the 8-atom Nb$_3$Sn unit cell, (b) two different 16-atom supercells, (c) one 24-atom supercell, and (d) the cubic 64-atom supercell. Turning now to our phonon calculations, for the 8-atom cell and the 2$\times$1$\times$1 16-atom cell, we use a 64-atom supercell with a 2$\times$2$\times$2 k-point mesh. For the other 16-atom cell, we use a 128-atom supercell with a 2$\times$2$\times$2 k-point mesh. For the 24-atom cell, we use a 96-atom supercell with a 2$\times$2$\times$3 k-point mesh. And for the 64-atom cell, we use the unit cell with a 2$\times$2$\times$2 k-point mesh.

\section{ISOLATED ANTISITE DEFECTS}

Antisite defects are the most common point defects in Nb$_3$Sn, with formation energies much lower than either species of vacancy \cite{Diffusion}. Their presence not only accounts for off-stoichiometry in Nb$_3$Sn, but also is the primary form of disorder in stoichiometric Nb$_3$Sn. In this paper, we will use Nb$_{Sn}$ and Sn$_{Nb}$ to refer to niobium atoms occupying tin sites, and tin atoms occupying niobium sites, respectively.

To calculate antisite defect concentration in stoichiometric Nb$_3$Sn, we consider the antisite defect pair formation energy. We calculate the pair formation energy at 0~K and free energy at the typical Nb$_3$Sn growth temperature of 1420~K, as Table 1 summarizes. We find that the pair formation free energy at 1420~K is nearly twice the pair formation energy at 0~K; this difference is attributable to
the decrease in electronic entropy associated with the defect pair, which in turn is a result of the negative effect of antisite defects on the density of states at the Fermi-level.  Specifically, such defects spread out the narrow peak in the density of states that appears at the Fermi-level in the undefected material (Fig. 1).

\begin{table}[h]
    \centering
    \setlength\tabcolsep{3.8pt}
    \caption{Antisite Defect Formation Free Energies}
    \begin{tabular}{@{}llcc@{}}
        \toprule
        \textbf{Defect} & \textbf{0~K Energy} & \textbf{1420~K Free Energy} & \textbf{Difference}\\
        \midrule
          Nb$_{Sn}$ & 0.31~eV & 0.51~eV & +0.20~eV\\
         \midrule
          Sn$_{Nb}$ & 0.31~eV & 0.56~eV & +0.25~eV\\
          \midrule
          Pair & 0.62~eV & 1.07~eV & +0.45~eV\\
        \bottomrule
    \end{tabular}
\end{table}

\begin{figure}
   \centering
   \includegraphics*[width=1\columnwidth]{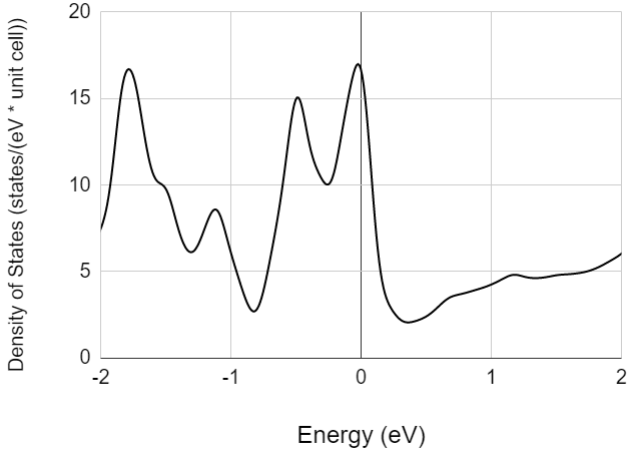}
   \caption{Calculated density of states for pure Nb$_3$Sn. The high peak at the Fermi-level plays a crucial role in determining many material properties.}
\end{figure}

We estimate that electronic entropy grows linearly with temperature due to the linear dependence of the width of the Fermi function on temperature. Therefore, we assume that the defect formation free energy is of the form $F = E-T\cdot(\alpha\cdot T)$, where we determine $\alpha$ based on our results in Table 1. We then add the configurational entropy contribution~$-T\cdot S_{config}(c)$ and minimize with respect to defect concentration find the equilibrium antisite defect concentration in stoichiometric Nb$_3$Sn as a function of temperature (Fig. 2). We note that, before accounting for electronic entropy effects, the predicted fractional antisite defect concentration is both very high (on the order of 0.1 at the coating temperature) and strongly temperature dependent. After properly accounting for electronic entropy effects, the predicted antisite defect concentration is an order of magnitude lower (about 0.01), and nearly constant above 1300 Kelvin. This is because the electronic entropy contribution to the free energy of antisite defects grows quadratically with $T$, while the configurational entropy contribution grows only linearly with $T$.

\begin{figure}
   \centering
   \includegraphics*[width=1\columnwidth]{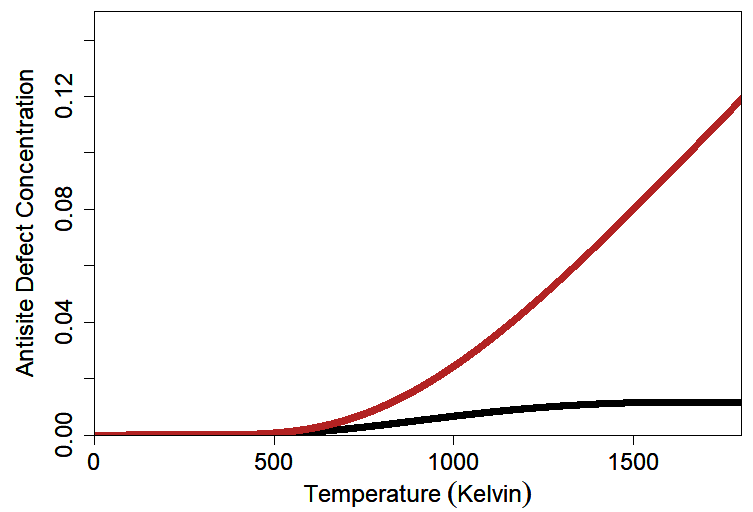}
   \caption{Calculated antisite defect concentration in 25\% Sn Nb$_3$Sn as a function of temperature, assuming defect formation energies that have been corrected (Black) and have not been corrected (Red) for electronic entropy effects.}
\end{figure}

An antisite defect concentration of 0.01 is not sufficiently high to significantly affect the electronic structure of Nb$_3$Sn, but it likely sets the electron mean-free path and, in general, may play a role in determining the amplitudes of important scattering processes. This is in contrast to the case of niobium, in which the electron mean-free path is determined by interstitial impurity concentrations.

The effect of antisite defects on the Fermi-level density of states not only determines the equilibrium defect concentration, but also determines how these defects interact with each other and with other defects. Defects that suppress the Fermi-level density of states in some volume of radius $R$ (visualized in Fig. 3) reduce the amount by which other defects can reduce the electronic entropy when they are present in that region.  We thus expect to find significant temperature-dependent attractive interaction between defects that reduce the Fermi-level density of states.

\begin{figure}
   \centering
   \includegraphics*[width=1\columnwidth]{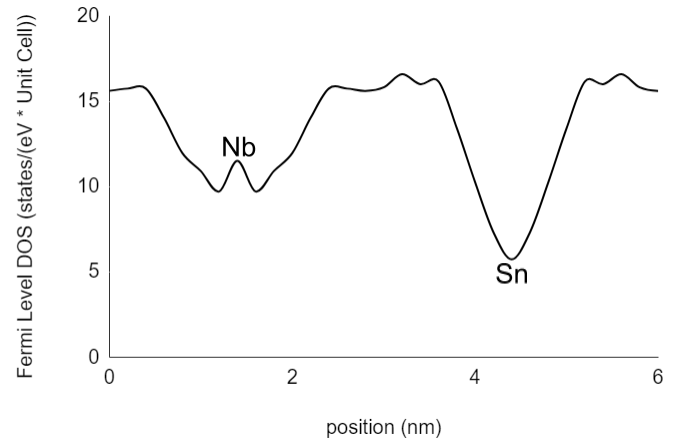}
   \caption{Real-space change in Fermi-level density of states due to niobium and tin antisite defects. We calculate Fermi-level density of states in slices in a 1x1x12 supercell with one of each antisite defect.}
\end{figure}

The above electronic entropy effect means that interactions between antisite defects, which are strongly repulsive at low temperatures, are only mildly repulsive at high temperatures. Understanding the temperature dependence of these interactions makes it possible to accurately calculate the stoichiometry limits of the A15 phase at high temperatures. Another possible implication of this electronic entropy effect is that antisite defects could be attracted to grain boundaries and dislocations at high temperatures, where the Fermi-level density of states may be lower due to disorder and strain. This could explain the existence of tin-rich regions around grain boundaries.

\section{PHASE DIAGRAM}

The appearance of compact tin-depleted regions in vapor-grown Nb$_3$Sn films\cite{Becker} is somewhat puzzling and perhaps suggestive of a tin-poor phase which may have been missed in past experimental studies. To explore this possibility, we calculate the A15 region of the Nb-Sn binary phase diagram, and find our results to be in good agreement with the experimentally observed phase diagram. We therefore propose a kinetic mechanism to explain the formation of tin-depleted regions and their distribution in the Nb$_3$Sn layer.

In outline, our procedure is as follows. First, we estimate equilibrium concentrations of the two species of antisite defects as functions of Nb-Sn relative chemical potential. For Nb antisite defects, we use the Monte Carlo method with a pairwise interaction model to estimate equilibrium concentration in a $8x8x8$ Nb$_3$Sn supercell. For Sn antisite defects, we estimate equilibrium concentration based on a Boltzmann factor approximation. Then, we take the resulting free energy versus stoichiometry data and use the convex hull approach to determine the final boundaries of the A15 phase and the competing bcc and Nb$_6$Sn$_5$ phases.

\subsection{Methodology}

To model the interactions among defects, we employ a cluster expansion at the pair-interaction level. Specifically, we consider all possible antisite defect pair interactions: $Nb_{Sn}$-$Nb_{Sn}$, $Nb_{Sn}$-$Sn_{Nb}$ and $Sn_{Nb}$-$Sn_{Nb}$. In order to make a realistic model that is as simple as possible, we neglect interactions that satisfy two criteria: (1) the relevant defects are dilute (more specifically the product of their concentrations is sufficiently small) and (2) the interactions are weak or repulsive (so that defect complexes are not a consideration).

Based on these criteria and preliminary interaction-energy calculations, we neglect $Nb_{Sn}$-$Sn_{Nb}$ and $Sn_{Nb}$-$Sn_{Nb}$ interactions, and therefore describe the concentration of $Sn_{Nb}$ defects with a simple Boltzmann factor approximation. We determine that the $Nb_{Sn}$-$Nb_{Sn}$ interaction is important, and calculate nearest, next-nearest, and next-next-nearest neighbor interactions to high precision (Table 2). We then use the Monte Carlo method to determine $Nb_{Sn}$ defect concentration as a function of temperature and as a function of Nb-Sn relative chemical potential. The Monte Carlo calculations were performed on a periodic 8x8x8 supercell of the 8-atom A15 unit cell using the Metropolis-Hastings algorithm, where a site is chosen at random and its occupancy is flipped according to the Boltzmann acceptance criterion.

\begin{table}
    \centering
    \setlength\tabcolsep{3.8pt}
    \caption{$Nb_{Sn}$ Pair Interaction Energies}
    \begin{tabular}{@{}llcc@{}}
        \toprule
        \textbf{Interaction} & \textbf{0~K Energy} & \textbf{1420~K Free Energy}\\
        \midrule
         Nearest-neighbor & 0.17~eV & 0.10~eV \\
         \midrule
         2nd Nearest-neighbor & 0.09~eV & 0.07~eV \\
         \midrule
         3rd Nearest-neighbor & 0.02~eV & 0.01~eV \\
        \bottomrule
    \end{tabular}
\end{table}

For our convex hull calculations, we consider the bcc niobium phase with tin substitutional defects and the Nb$_6$Sn$_5$ phase as the competing phases on the tin-poor and tin-rich A15 boundaries, respectively (Fig. 4). While Nb$_6$Sn$_5$ is experimentally observed only below temperatures of about 1200 K, we use it as a reference at all temperatures due to the difficulty in precisely calculating the free energy of the liquid solution that competes with the A15 phase at higher temperatures. Therefore, our calculated tin-rich stoichiometry limit should be considered an upper bound for the true limit at temperatures above 1200 K.

\begin{figure}
   \centering
   \includegraphics*[width=1\columnwidth]{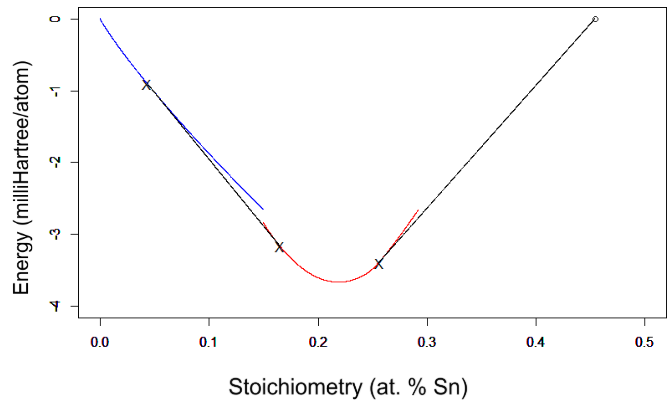}
   \caption{Example convex hull calculation, at $T$=790~K. Linear tangent lines connect the convex hulls of the bcc and A15 phases and the point energy calculated for Nb$_6$Sn$_5$. The tangent points, marked with X's, indicate the calculated stoichiometry limits of the phases.}
\end{figure}

\subsection{Results}

We accurately reproduce the A15 region of the experimentally determined Nb-Sn phase diagram. In particular, we find a lower limit of $16.5\%$ tin content, and an upper limit of $27.1\%$ tin content at 1420~K, close to the experimental values of about 17.5\% and 26\% respectively (Fig. 5). We also note the smooth, convex nature of the A15 hull, which shows no signs of separating into multiple A15 phases, leaving the presence of distinct tin-depleted regions in Nb$_3$Sn thin films somewhat of a mystery to be discussed further below. 

\begin{figure}
   \centering
   \includegraphics*[width=1\columnwidth]{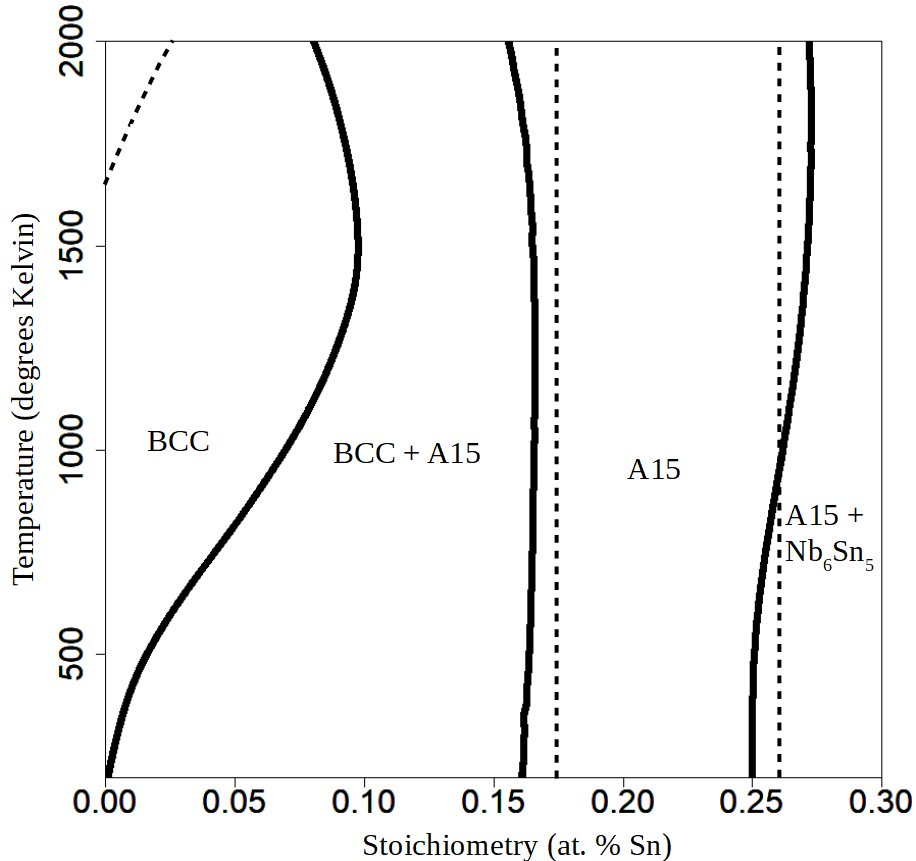}
   \caption{Calculated Nb-Sn Phase Diagram. The experimentally established phase diagram is represented by dashed lines for comparison\cite{Charlesworth}.}
\end{figure}

\subsection{Discussion}

The tin-rich limit of the A15 phase is found to rise smoothly from 25\% at low temperatures to a maximum just above 27\% at high temperatures. The low-temperature limit of this phase boundary is of high confidence; therefore, we propose a minor adjustment to the established phase diagram: instead of a constant tin-rich stoichiometry limit of 26\%, the tin-rich limit should smoothly approach 25\% at low temperatures \cite{Charlesworth}. The calculated tin-rich limit at higher temperatures is somewhat higher than the experimentally observed tin-rich limit. We believe this is due to our use of Nb$_6$Sn$_5$ as the competing phase at high temperatures, when in reality it is known that the Sn-Nb liquid solution phase has a lower free energy beyond about 1200~K. Accounting for this other phase would lower our calculated tin-rich stoichiometry limit, bringing it in closer agreement with experimental values, but to do so accurately is beyond the scope of this work.

The tin-poor limit of the A15 phase is found to lie between 16\% and 17\% Sn throughout the calculated temperature range, in good agreement with most experimental studies. As might be expected, our Monte Carlo simulations find that Nb antisite defects begin to fall into low-energy superlattice configurations at low temperatures. Therefore, in this regime, an accurate calculation of the phase boundary requires determination of the minimum-energy superlattice configuration for Nb antisite defects. Triplet interactions and other corrections to our Monte Carlo model may be necessary in order to make this determination. As a result, our calculated phase boundary is of slightly lower confidence at low temperatures. We find no evidence, however, for the narrowing of the stoichiometry range at low temperatures suggested by some researchers \cite{Devantay}.

The tin solubility limit of the bcc Nb phase is quite uncertain experimentally, with most measurements below $\sim$1500K finding no more than $\sim$1\% solubility, but at least one indirect measurement finding much higher solubility on the order of 10\% \cite{Schiffman}. Our calculation is in agreement with the latter, suggesting that the phase boundary passes 1\% Sn by 500K, and increases to a maximum of over 10\% near the coating temperature. It is possible that some experiments failed to access the phase boundary due to the slow kinetics of Sn diffusion in Nb and Nb$_3$Sn. Self-diffusion studies of Nb found that it would take about a day to achieve self-diffusion across 100~nm at the coating temperature of 1420K, or about a month to achieve self-diffusion across 100~nm at 1300K \cite{DiffusionNb}. It is also possible that a systematic error in DFT-calculated energy differences between the A15 and bcc phases results in unrealistic tin solubility estimates in our model. An in-depth investigation of this problem is beyond the scope of this paper, but we note that the calculated tin-poor limit of the A15 phase is not very sensitive to changes in the tin-rich limit of the bcc phase.

Our calculations predict that the A15 phase should not phase separate, which improves our understanding of the experimentally observed tin-depleted regions in Nb$_3$Sn samples. We conclude that this inhomogeneity is likely due to kinetic phenomena, not due to equilibrium phase separation. For kinetics to be the determining factor, two things must be true: (a) the diffusion of tin and/or niobium within the A15 phase must be extremely slow at the growth temperature and (b) the chemical potential must change abruptly at certain times during layer growth.

The slow diffusion of tin in the A15 phase has been described in detail in a first-principles study by \cite{Diffusion}. To summarize, the activation energy for diffusion on tin sites is the sum of the niobium vacancy formation energy, the niobium vacancy hopping energy, and the antisite energy, a quantity of about 4~eV. This high activation energy results in an extremely low diffusion rate. As a result, it is a good approximation to say that the stoichiometry of a tin-poor A15 crystal is essentially static at the typical growth temperature, except in regions within a few nanometers of a diffusion pathway, such as an interface or a grain boundary.

To explain the abrupt change in chemical potential, we hypothesize that there are two distinct growth states for the Nb$_3$Sn layer in the vapor diffusion setup. In the tin-rich growth state, tin vapor arrives on the surface at a faster rate than it is consumed in the Nb$_3$Sn formation reaction, resulting in the accumulation of a layer of liquid tin on the surface. This then results in a high chemical potential for Sn, and the Nb$_3$Sn then grows at a tin-rich stoichiometry. In the tin-poor growth state, tin vapor arrives on the surface at a slower rate than it is consumed in the Nb$_3$Sn formation reaction, resulting in the depletion of any excess surface tin. This results in a low tin chemical potential and Nb$_3$Sn growth at a tin-poor stoichiometry. Specifically, we expect an abrupt change from tin-rich to tin-poor growth at times when the amount of excess tin on the surface drops to zero.

\begin{figure}
   \centering
   \includegraphics*[width=1\columnwidth]{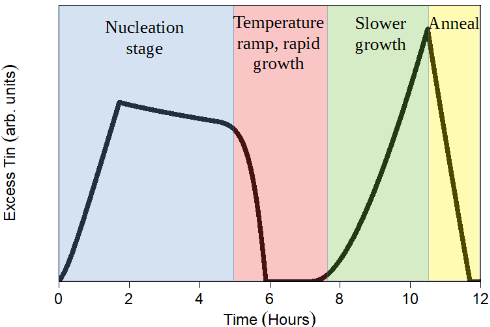}
   \caption{Estimated excess tin on the cavity surface as a function of time throughout the coating process, using a simple model that accounts for the temperature-dependent evaporation rates of tin and tin chloride, and the thickness-dependent growth rate of the Nb$_3$Sn layer. We expect tin-poor growth at times when there is no excess surface tin.}
\end{figure}

These considerations lead to the following understanding of the composition of vapor-deposition grown Nb$_3$Sn. Based on experimental measurements, we estimate the amount of excess tin on the Nb$_3$Sn surface as a function of time during the growth process (Fig. 6). A tin excess is established early on, when the surface is seeded with a tin chloride nucleation agent at low temperatures, and diffusion and reaction rates are slow \cite{DHall}. This ensures a tin-rich growth state early on, which is very important because this early growth ultimately constitutes the surface of the finished Nb$_3$Sn layer. When this initial excess of tin has been exhausted, we hypothesize that tin-poor stoichiometry growth begins, and regions of tin-poor stoichiometry form. At later times, Nb$_3$Sn growth slows down as the layer thickens, some excess tin again accumulates on the surface, and tin-rich growth resumes. Finally, during the annealing step, tin vapor arrival is halted and the remaining surface tin is consumed, resulting in another period of tin-poor growth. Therefore, we expect the layer to be characteristically tin-rich near the surface and tin-poor near the Nb-Nb$_3$Sn interface, with regions of tin-poor stoichiometry in the bulk of the layer, consistent with the experimental findings.

\section{T$_c$ VERSUS STOICHIOMETRY STUDY AND GRAIN BOUNDARIES}

The critical temperature of A15 Nb-Sn has been studied experimentally, with the conclusion that $T_c$ has a maximum value of about 18K near the 25\% Sn stoichiometry of Nb$_3$Sn, and a minimum value of about 6K at the tin-poor stoichiometry limit of 17-18\% Sn \cite{Tc}. However, the superconducting properties of tin-rich stoichiometries beyond 26\% Sn recently observed in around grain boundaries have not been measured \cite{GB}. Motivated to understand the effect of these tin-rich grain boundaries on Nb$_3$Sn cavity performance, we perform the first calculation of T$_c$ as a function of stoichiometry in A15 Nb-Sn.

We use DFT to calculate $T_c$ using Eliashberg theory \cite{Eliashberg}. To do this, we calculate electron-phonon scattering amplitudes in A15 cells across the observed stoichiometry range. In order to make these calculations as precise as possible, we employ Wannier function methods to integrate smoothly over all scattering processes in momentum space \cite{mlwf,FeynWann1,FeynWann2}. These scattering amplitudes are used to calculate the phonon spectral function (Fig. 7), which can be used to estimate $T_c$ using the McMillan formula \cite{McMillan}.

A previous first-principles study of Nb$_3$Sn by Mentink et al. explores the effect of electron-lifetime broadening on superconducting properties, and finds that greater electron-lifetime broadening results in a lower Fermi-level density of states, and a lower T$_c$  \cite{Mentink}. In these calculations, we used a fixed broadening energy of 0.01 eV for electronic states, on the lower end of the range explored by Mentink and consistent with low normal state resistivity Nb$_3$Sn. Following Mentink et al., we use an energy broadening of 0.0012 eV for phonon states, and a constant effective coulomb repulsion term $\mu^* = 0.125$.

\begin{figure}
   \centering
   \includegraphics*[width=1\columnwidth]{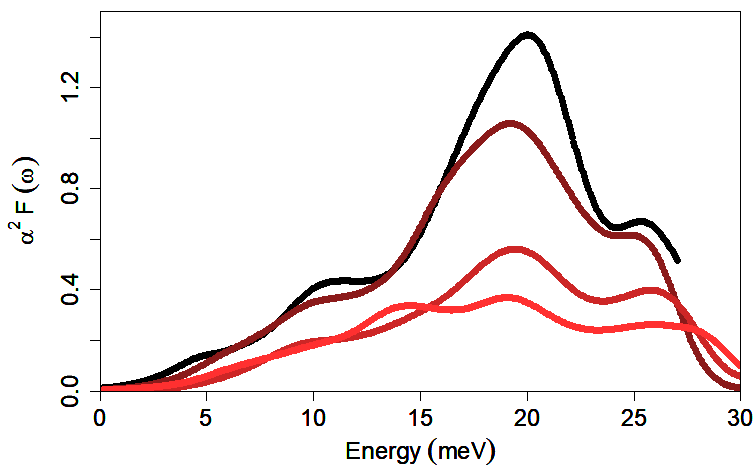}
   \caption{Example Phonon spectral function calculations for Nb$_3$Sn (black), 26.6\% (dark red), 29.2\% (red), and 31.3\% (light red) A15 cells. The calculated $\alpha ^2F(\omega)$ decreases at all omega with increasing tin content above 25\%.}
\end{figure}

Our calculated $T_c$ values are in good agreement with experiment for calculations near 25\% Sn stoichiometry, and modestly above experiment for calculations at tin-poor stoichiometries (Table 3). A possible explanation for this discrepancy is that we did not account for possible changes in electron lifetime broadening. Tin-poor stoichiometry samples are likely to be more disordered than near-25\% samples, so a larger broadening energy may be appropriate for calculations on tin-poor stoichiometries. The results of Mentink et al. suggest that this adjustment would likely lead to lower calculated T$_c$ values \cite{Mentink}.

For experimentally inaccessible tin-rich stoichiometries, we find that $T_c$ falls to a minimum of about 5K at 31.25\% Sn (Fig. 8). This information, taken together with experimental measurements of the stoichiometry profile around grain boundaries from the Seidman group \cite{GB}, has enabled simulations of magnetic flux entry at grain boundaries performed by the Transtrum group (Fig. 9). These simulations show that because the coherence length of Nb$_3$Sn is on the order of a few nanometers, similar to the size of the tin segregation region, tin-rich grain boundaries can admit flux vortices even at modest fields, resulting in degraded cavity performance \cite{Transtrum1,Transtrum2}.

\begin{figure}
   \centering
   \includegraphics*[width=1\columnwidth]{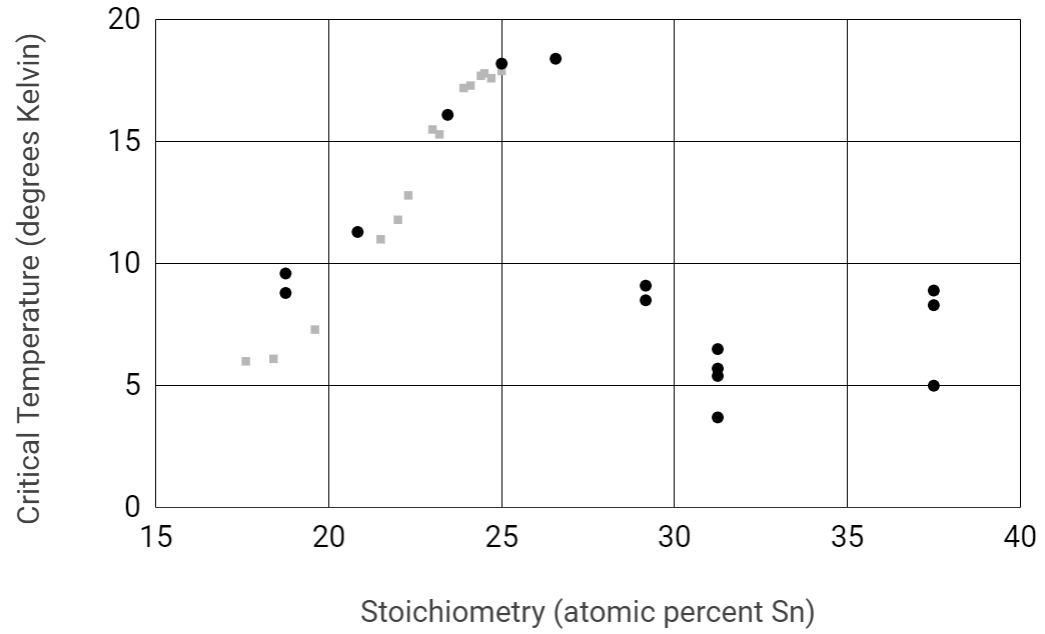}
   \caption{Experimental $T_c$ \cite{Devantay} (grey squares) and calculated $T_c$ (black circles) for A15 Nb-Sn of different stoichiometries. The calculated $T_c$ reaches a minimum of about 5 Kelvin in the tin-rich regime.}
   \label{fig:Tc}
\end{figure}

\begin{figure}
   \centering
   \includegraphics*[width=1\columnwidth]{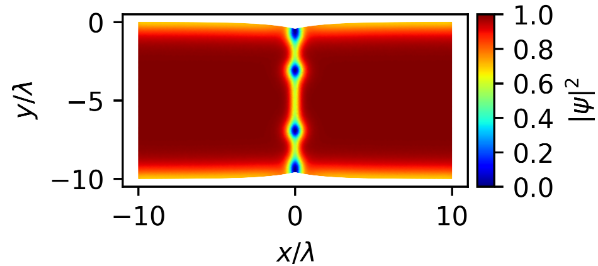}
   \caption{Example simulation of the superconducting order parameter in a tin-rich grain boundary, showing vortex penetration. The axes are in units of the RF penetration depth: $\lambda \approx$ 100~nm.}
\end{figure}

\begin{table}
    \centering
    \setlength\tabcolsep{3.8pt}
    \caption{Calculated vs. Measured $T_c$}
    \begin{tabular}{@{}llcc@{}}
        \toprule
        \textbf{Composition} & \textbf{Experimental T$_c$ (K)} & \textbf{Calculated T$_c$ (K)} \\
         \midrule
          18.75\% Sn & 6 & 9.2$^{\dagger}$ \\
         \midrule
          20.83\% Sn & 9.5 & 11.3  \\
         \midrule
          23.44\% Sn & 16 & 16.1  \\
         \midrule
          Nb$_3$Sn & 18 & 18.2 \\
        \bottomrule
    \end{tabular}
    $\dagger$ Averaged over two configurations.
\end{table}

Additionally, we find a strong correlation between $T_c$ and the Fermi-level density of states in our calculations for off-stoichiometric unit cells (Fig. 10). If this relationship is true in general for the A15 phase, it could be a useful tool in ongoing efforts to estimate superconducting properties of grain boundaries as an alternative to preforming full Eliashberg theory calculations which become very computationally expensive for complex defects such as grain boundaries.
\begin{figure}
   \centering
   \includegraphics*[width=1\columnwidth]{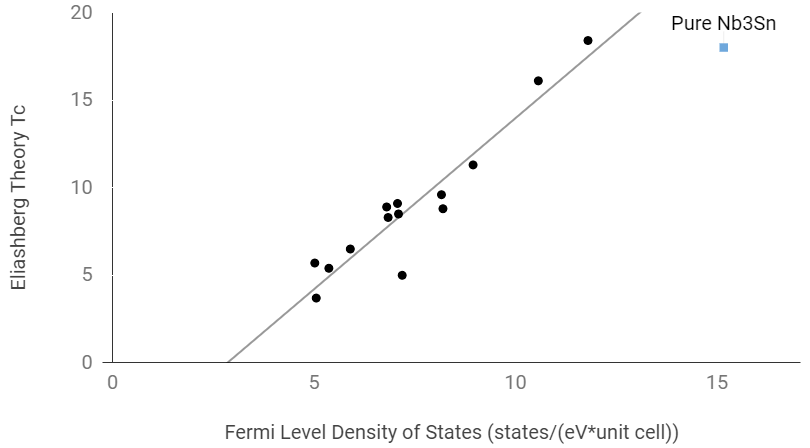}
   \caption{Correlation between $T_c$ and Fermi-level density of states in A15 cells of different stoichiometries. Calculated $T_c$ for off-stoichiometric A15 cells has an approximately linear relationship with Fermi-level density of states.}
\end{figure}

Finally, we address the question of why some grain boundaries have tin-rich stoichiometries in the first place, an interesting one in its own right. We propose a simple model for grain boundaries that assumes the Fermi-level density of states is reduced due to disorder and strain in some region around the grain boundary, and that the strong temperature dependence of antisite defect formation free energies is reduced proportionately. This would allow the A15 phase to reach higher tin stoichiometries in a small region around the grain boundary, as is seen experimentally. Our calculations show that a modest reduction in the electronic free energy contribution to the antisite energy would allow the tin-rich limit of the A15 phase to extend beyond 30\% tin stoichiometry (Fig. 11), similar to what is seen experimentally.

\begin{figure}
   \centering
   \includegraphics*[width=1\columnwidth]{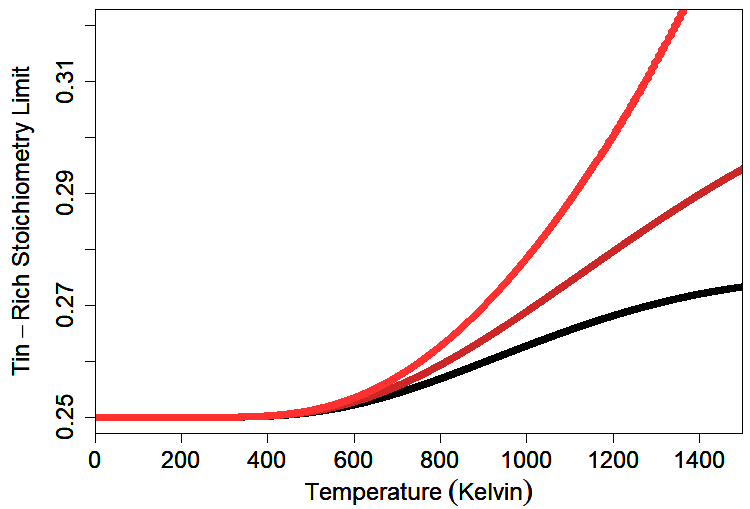}
   \caption{Comparison between the calculated tin-rich stoichiometry limit of the A15 phase with the temperature dependence of the tin antisite energy as calculated (black), reduced by 25\% (dark red), and reduced by 50\% (light red).}
\end{figure}

\section{CONCLUSIONS}

Our DFT study of antisite defects has given us a valuable new perspective on their thermodynamic behavior during layer growth and on their influence on the SRF properties of Nb$_3$Sn cavities. In particular, based on our understanding of how Nb antisite defects cluster together to form tin-depleted regions and how Sn antisite defects at grain boundaries affect their superconducting properties, we can more confidently link parameters of the Nb$_3$Sn growth process to cavity performance.

It has long been known that the superconducting performance of the A15 Nb-Sn phase is maximized in the vicinity of perfect Nb$_3$Sn stoichiometry, and therefore that tin-poor stoichiometry is a threat to the performance of a Nb$_3$Sn superconductor. When researchers found that tin-depleted regions do indeed exist in Nb$_3$Sn cavities, it quickly raised questions of how, why, and where they form. Our results provide new evidence that they do not form due to the diffusion of Nb antisite defects into low-energy clusters, but rather due to changes in chemical potential during layer growth. Given this, we believe that tin-depleted regions do not form at random, but instead form at predictable depths that can be controlled by changing coating parameters. This understanding makes it possible to quickly assess whether any potential Nb$_3$Sn growth protocol is likely to result in tin-depleted regions near the surface, where they are most likely to affect cavity performance.

When researchers found tin segregation at grain boundaries, it raised similar questions about how and why Sn antisite defects segregate at grain boundaries, and the new question of what effect, if any, tin-rich stoichiometry has on superconducting properties. Our calculations shed light on both of these topics; they offer an explanation for why the tin-rich stoichiometry limit of the phase extends well above the usual 26\% in the vicinity of grain boundaries, and also indicate that tin-rich stoichiometry results in degraded $T_c$. Both of these effects are directly attributable to the negative effect of Sn antisite defects on the Fermi-level density of states. This finding supports the hypothesis that it is important to anneal Nb$_3$Sn for a long enough time that the Sn chemical potential falls and grain boundaries to lose their excess Sn (but not such a long time that longer-range diffusion occurs and the layer becomes tin-depleted).

In the future, we plan to use DFT to study the behavior of antisite defects and grain boundaries in even greater detail, building a quantitative link between the composition of Nb$_3$Sn layers and the SRF performance of Nb$_3$Sn cavities. We hope these findings will inspire new Nb$_3$Sn growth protocols that push cavity performance to new heights.

\section{acknowledgment}

We would like to thank Michelle Kelley of Cornell University for many useful conversations about this research and help editing this manuscript, Jae-yel Lee and David Seidman of Northwestern University for sharing their grain boundary segregation measurements with us, and Sam Posen of Fermilab for organizing, facilitating, and contributing valuable insights to the many meetings we had with Dr. Lee and Dr. Seidman.

This work was supported by the US National Science Foundation under award PHY-1549132, the Center for Bright Beams.


\begin{thebibliography}{99} 

\bibitem{Posen}
    S. Posen, M. Liepe. D. Hall,
    ``Proof-of-principle demonstration of Nb3Sn superconducting RF cavities for high Q0 applications,'' Applied Physics Letters 106, Issue 8, February 2015,
    dx.doi.org/10.1063/1.4913247

\bibitem{Liepe}
    S. Posen and M. Liepe,
    ``Advances in development of Nb3Sn superconducting radio-frequency cavities,''
    Phys. Rev. ST Accel. Beams 15, 112001, 2014, journals.aps.org/prstab/abstract/10.1103/PhysRevSTAB.17.112001

\bibitem{Ryan}
    R. D. Porter, T. A. Arias, P. D. Cueva, M. Liepe, D. A. Muller, A. R. Pack, N. Sitaraman, M. K. Transtrum,
    ``Field Limitation in Nb3Sn Cavities, 19th International Conference on RF Superconductivity,''
    Proc. SRF'19, Dresden, Germany, July 2019,
    Paper THFUA5
 
\bibitem{DHall}
    D. L. Hall,
    ``New Insights into the Limitations on the Efficiency and Achievable Gradients in Nb$_3$Sn SRF Cavities,''
    PhD. Thesis, Physics Department, 2017

\bibitem{LINAC}
    R. D. Porter, T. Arias, P. Cueva, D. L. Hall, M. Liepe, J. T. Maniscalco, D. A. Muller, N. Sitaraman,
    ``Next Generation Nb3Sn Cavities for Linear Accelerators,''
    Proc. LINAC’18, Beijing, China 2018,
    Paper TUP055
    
\bibitem{Lee}
    J. Lee, S. Posen, Z. Mao, Y. Trenikhina, K. He, D. L. Hall, M. Liepe, D. N. Seidman,
    ``Atomic-scale analyses of Nb3Sn on Nb prepared by vapor diffusion for superconducting radiofrequency cavity applications: a correlative study,''
    Superconductor Science and Technology, 32, 2, 024001, 2018

\bibitem{Charlesworth}
    J. P. Charlesworth, I. Macphail, and P. E. Madsen,
    ``Experimental work on the niobium-tin constitution diagram and related studies,''
    J Mater Sci (1970) 5: 580, 1970,
    doi.org/10.1007/BF00554367

\bibitem{GB}
     J. Lee, Z. Mao, K. He, Z. H. Sung, T. Spina, S.-I. Baik, D. L. Hall, M. Liepe, D. N. Seidman,  and S. Posen, arXivpreprint arXiv:1907.00476  (2019)
    ``Grain-boundary segregation behavior in Nb3Sn coatings on Nb for superconducting radiofrequency cavity applications,''
    arXiv:1907.00476  (2019)

\bibitem{Diffusion}
    R. Besson, S. Guyot, and A. Legris,
    ``Atomic-scale study of diffusion in A15 Nb$_3$Sn,''
    Phys. Rev. B 75, 054105, February 2007,
    journals.aps.org/prb/abstract/10.1103/PhysRevB.75.054105

\bibitem{Coherence}
    D. B. Liarte, S. Posen, M. K. Transtrum, G. Catelani, M. Liepe, and J. P. Sethna,
    ``Theoretical estimates of maximum fields in superconducting resonant radio frequency cavities: stability theory, disorder, and laminates,''
    Superconductor Science and Technology 30, 033002, 2017

\bibitem{JDFTx}
	R. Sundararaman, K. Letchworth-Weaver, K. A. Schwarz, D. Gunceler, Y. Ozhabes and T. A. Arias,
	``JDFTx: software for joint density-functional theory,''
	SoftwareX 6, 278, 2017

\bibitem{GGA}
    J. P. Perdew, K. Burke and M. Ernzerhof,
    ``Generalized Gradient Approximation Made Simple,''
    Phys. Rev. Lett. 77, 3865, 1996,
    journals.aps.org/prl/abstract/10.1103/PhysRevLett.77.3865

\bibitem{Psp}
    K. F. Garrity, J. W. Bennett, K. M. Rabe and D. Vanderbilt,
    ``Pseudopotentials for high-throughput DFT calculations,''
    Comput. Mater. Sci. 81, 446, 2014,
    doi.org/10.1016/j.commatsci.2013.08.053

\bibitem{Cold}
    N. Marzari, D. Vanderbilt, A. De Vita, and M. C. Payne,
    ``Thermal Contraction and Disordering of the Al(110) Surface,''
    Phys. Rev. Lett., 82:3296, 1999,
    journals.aps.org/prl/abstract/10.1103/PhysRevLett.82.3296

\bibitem{Becker}
    C. Becker, S. Posen, N. Groll, R. Cook, C. M. Schlepütz, D. L. Hall, M. Liepe, M. Pellin, J. Zasadzinski, and T. Proslier,
    ``Analysis of Nb3Sn surface layers for superconducting radio frequency cavity applications,''
    Appl. Phys. Lett. 106 082602, 2015

\bibitem{Tc}
    A. Godeke,
    ``A review of the properties of Nb3Sn and their variation with A15 composition, morphology and strain state,''
    Superconductor Science and Technology, 19(8):R68—-R80, August 2006

\bibitem{Devantay}
    H. Devantay, J. L. Jorda, M. Decroux, J. Muller, and R. Flükiger,
    ``The physical and structural properties of superconducting A15-type Nb-Sn alloys,''
    J Mater Sci, 16: 2145, 1981,
    doi.org/10.1007/BF00542375

\bibitem{Schiffman}
    R. A. Schiffman and D. M. Bailey,
    ``Thermodynamics of the Incongruently Subliming Niobium-Tin System,''
    HighTemp. Sci., 15, pp. 165-77, 1982

\bibitem{DiffusionNb}
    R. E. Einziger, J. N. Mundy, and H. A. Hoff,
    ``Niobium self-diffusion,''
    Phys. Rev. B 17, 2, August 1977,
\bibitem{Eliashberg}
    G. M. Eliashberg,
    ``Interactions between Electrons and Lattice Vibrations in a Superconductor,''
    Zh. Eksperim. i Teor. Fiz. 38 966, 1960

\bibitem{mlwf}
    N. Marzari and D. Vanderbilt,
    ``Maximally localized generalized Wannier functions for composite energy bands,''
    Phys. Rev. B 56, 12847, November 1997
    journals.aps.org/prb/abstract/10.1103/PhysRevB.56.12847

\bibitem{FeynWann1}
    A. Brown, R. Sundararaman, P. Narang, W. A. Goddard III and H. A. Atwater, ``Non-Radiative Plasmon Decay and Hot Carrier Dynamics: Effects of Phonons, Surfaces and Geometry,''
    ACS Nano, 10, 957, 2016
    
\bibitem{FeynWann2}
    A. Habib, F. Florio and R. Sundararaman,
    ``Hot carrier dynamics in plasmonic transition metal nitrides,''
    J. Opt. 20, 064001, 2018
    
\bibitem{McMillan}
    W. L. McMillan,
    ``Transition Temperature of Strong-Coupled Superconductors,''
    Phys. Rev. 167, 331, March 1968,
    journals.aps.org/pr/abstract/10.1103/PhysRev.167.331

\bibitem{Mentink}
    M. G. T. Mentink, M. M. J. Dhalle, D. R. Dietderich, A. Godeke, F. Hellman, and H. H. J. ten Kate,
    ``The effects of disorder on the normal state and superconducting properties of Nb3Sn,''
    Supercond. Sci. Technol. 30, 025006, 2017,
    doi.org/10.1088/1361-6668/30/2/025006
\bibitem{Transtrum1}
    A. R. Pack, J. Carlson, S. Wadsworth, and M. K. Transtrum, 
    ``Role of inhomogeneities for vortex nucleation in superconductors within time-dependent Ginzburg-Landau theory,''
    arXiv preprint arXiv:1911.02132, November 2019
\bibitem{Transtrum2}
    J. Carlson, A. Pack, M. K. Transtrum, S. Posen, J. Lee, D. N. Seidman, D. B. Liarte, N. Sitaraman, A. Senanian, T. A. Arias, and J. P. Sethna,
    ``Analysis of Magnetic Vortex Dissipation in Sn Segregated Boundaries in Nb3Sn SRF Cavities,''
    to be submitted as arXiv preprint, December 2019
\end{thebibliography}
\end{document}